\PassOptionsToPackage{hyphens}{url}
\documentclass[10pt]{article}

\usepackage{xcolor}
\usepackage{amsmath,amssymb}
\usepackage{lmodern}

\usepackage[T1]{fontenc}
\usepackage[utf8]{inputenc}
\usepackage{textcomp}

\IfFileExists{upquote.sty}{\usepackage{upquote}}{}
\IfFileExists{microtype.sty}{%
  \usepackage[]{microtype}
  \UseMicrotypeSet[protrusion]{basicmath} 
}{}
\makeatletter
\@ifundefined{KOMAClassName}{
  \IfFileExists{parskip.sty}{%
    \usepackage{parskip}
  }{
    \setlength{\parindent}{0pt}
    \setlength{\parskip}{6pt plus 2pt minus 1pt}}
}{
  \KOMAoptions{parskip=half}}
\makeatother

\usepackage{longtable,booktabs,array,calc}
\usepackage{etoolbox}
\IfFileExists{footnotehyper.sty}{\usepackage{footnotehyper}}{\usepackage{footnote}}
\makesavenoteenv{longtable}

\usepackage{graphicx}
\graphicspath{{figures/}{figures/media/}}
\makeatletter
\newsavebox\pandoc@box
\newcommand*\pandocbounded[1]{
  \sbox\pandoc@box{#1}%
  \Gscale@div\@tempa{\textheight}{\dimexpr\ht\pandoc@box+\dp\pandoc@box\relax}%
  \Gscale@div\@tempb{\linewidth}{\wd\pandoc@box}%
  \ifdim\@tempb\p@<\@tempa\p@\let\@tempa\@tempb\fi
  \ifdim\@tempa\p@<\p@\scalebox{\@tempa}{\usebox\pandoc@box}%
  \else\usebox{\pandoc@box}%
  \fi%
}
\def\fps@figure{htbp}
\makeatother

\usepackage{float}       
\usepackage[section]{placeins}

\usepackage{xurl} 
\usepackage{bookmark}
\usepackage{hyperref}
\hypersetup{
  hidelinks,
  pdfcreator={LaTeX via pandoc}
}
\usepackage{authblk}

\title{\emph{DevLicOps}: A Framework for Mitigating Licensing Risks in AI-Generated Code}

\author[1]{Pratyush Nidhi Sharma$^{*}$}
\author[2]{Lauren Wright}
\author[1]{Anne Herfurth}
\author[1]{Munsif Sokiyna}
\author[3]{Pratyaksh Nidhi Sharma}
\author[1]{Sethu Das}
\author[1]{Mikko Siponen}
\affil[1]{The University of Alabama, USA}
\affil[2]{Insitely Technology Solutions, USA}
\affil[3]{HRMNY HR, USA}

\date{}

\begin{document}
\maketitle
\begingroup
\renewcommand\thefootnote{}
\footnotetext{%
\textsuperscript{*}Corresponding author \quad
\textit{Email addresses}:
\texttt{pnsharma@ua.edu} (Pratyush Nidhi Sharma),\quad
\texttt{lauren@insitely.us} (Lauren Wright),\quad
\texttt{aherfurth@ua.edu} (Anne Herfurth),\quad
\texttt{msokiyna@crimson.ua.edu} (Munsif Sokiyna),\quad
\texttt{pratyaksh.sharma@hrmny-hr.com} (Pratyaksh Nidhi Sharma),\quad
\texttt{sdas11@crimson.ua.edu} (Sethu Das),\quad
\texttt{tmsiponen@ua.edu} (Mikko Siponen).
}
\addtocounter{footnote}{0} 
\endgroup

\begin{abstract}
Generative AI coding assistants (ACAs) are widely adopted yet pose serious legal and compliance risks. ACAs can generate code governed by restrictive open-source licenses (e.g., GPL), potentially exposing companies to litigation or forced open-sourcing. Few developers are trained in these risks, and legal standards vary globally, especially with outsourcing. Our article introduces DevLicOps, a practical framework that helps IT leaders manage ACA-related licensing risks through governance, incident response, and informed tradeoffs. As ACA adoption grows and legal frameworks evolve, proactive license compliance is essential for responsible, risk-aware software development in the AI era.
\end{abstract}


\section{Introduction}\label{introduction}

Generative AI coding assistants (ACAs) are revolutionizing how developers write code. A 2024 survey by GitHub reports that a staggering 97\% of developers working in large corporations in U.S., Brazil, Germany, and India now use ACAs while at work.\footnote{\url{https://github.blog/news-insights/research/survey-ai-wave-grows/}} Developers now have an increasingly wide array of ACAs to choose from {[}8{]}. However, only 30--40\% of respondents noted that their company is actively encouraging and promoting the adoption of ACAs. The remaining respondents, representing a significant proportion of companies, share that their company either prohibits the use of ACAs, is indifferent, or offers only limited encouragement. Another global study conducted by the cloud security firm Checkmarx found that only 29\% of surveyed companies had some form of AI related governance in place.\footnote{\url{https://checkmarx.com/7-steps-genai-survey-gen/}} A separate survey noted that 60\% of organizations lack formal processes for assessing ACA-generated code.\footnote{\url{https://www.harness.io/state-of-software-delivery}} Despite the many advantages offered by ACAs, such as efficiency and speed in coding, the gap between official AI policies and developers' actual practices creates a new frontier of license compliance-related legal and ethical risks for companies {[}1, 2, 3{]}. The core issue is that ACAs can generate code protected by restrictive open-source licenses. These risks extend beyond legal disputes and can jeopardize mergers, acquisitions, and other strategic deals by complicating business valuations {[}2{]}.

Consider these realistic scenarios:

\begin{enumerate}
\def\labelenumi{\arabic{enumi}.}
\item
  An IT manager at a mid-sized software firm receives a startling email that a regulatory audit has uncovered non-compliant open-source code within their product. An investigation reveals that large portions of code were generated by a developer using various ACAs. With a looming product release, the manager must quickly figure out how to address the violations, prevent future occurrences, and avoid potential lawsuits.
\item
  An IT manager at a small firm is reviewing the latest deliverables from an offshore software development partner. A casual comment from one of the offshore developers raises a red flag: they used ACAs to expedite the coding process without any open-source license compliance checks. The manager realizes that the company's contract lacks explicit clauses regarding the use of AI-generated code. With the offshore team potentially operating under different copyright and intellectual property laws, the manager now faces a daunting challenge: how to verify the compliance of the delivered code and prevent potential legal and reputational damage.
\item
  An IT manager's development team requests to use ACAs for efficient product development. However, a senior developer raises concerns about the AI-generated code's origins and potential licensing issues. The IT manager realizes that the team lacks a clear policy on how AI-generated code should be used or reviewed. The manager must now develop a policy to guide the team without stifling innovation.
\end{enumerate}

As ACAs become increasingly prevalent in coding workflows, scenarios like these are likely to unfold repeatedly across the globe. These are not abstract concerns but real and pressing threats that IT managers are facing. For instance, a senior IT manager with over 15 years of experience in a U.S. based technology firm expressed deep concern during our interview about ``license laundering'' and the provenance of AI-generated code. The manager also highlighted that their firm remains uncertain about how to shape an effective ACA policy due to developers' divergent views. In contrast, junior developers may not be aware of the implications of licensing issues when using ACAs. There may be substantial differences in how seasoned developers perceive these issues than less experienced developers. University computer science and IT curriculums often do not include legal and regulatory compliance training, and fresh developers joining the workforce may not have requisite knowledge of open-source licensing. As the new generation of software developers learns coding from and grows up reliant on ACAs, these issues will become increasingly prevalent in the foreseeable future. Additionally, awareness, perspectives, and regulations can vary substantially across the globe.

ACAs are typically trained on unsanitized open-source code from the internet and, as a result, may produce code as output that is protected by open-source licenses, including restrictive copyleft licenses like the GPL and AGPL {[}3, 4, 8{]}. Generative AI models are known to memorize, and sometimes regurgitate verbatim, data they were trained on, a phenomenon known as training data leakage {[}5, 6, 7{]}. The regurgitation rates have found to be as high as 80\% for some ACAs {[}8{]}. While small snippets of functional ACA-generated code are unlikely to pose issues, the likelihood of copyrightability concerns increases with larger and more complex code suggestions {[}3{]}.\footnote{\url{https://katedowninglaw.com/2023/07/12/copilot-and-snippet-scanning/}} This risk will continue to grow in the future as companies and non-developers increasingly explore vibe coding and deploying full-stack, end-to-end software using fully ACA-generated code.\footnote{\url{https://medium.com/@colinbaird_51123/tips-for-building-an-application-from-scratch-with-chatgpt-as-a-non-developer-9b256336d6aa}} Furthermore, a significant portion of ACA training is likely to be based on older open source code, some of which may now be governed under more restrictive copyleft licenses.\footnote{\url{https://eastwind.substack.com/p/the-future-of-programming-copilots}}\textsuperscript{,} \footnote{\url{https://sdtimes.com/os/navigating-unexpected-license-changes-in-open-source-software/}}

Incorporating such code could lead to licensing violations as highlighted by the highly-publicized class action lawsuit by open-source programmers against GitHub.\footnote{\url{https://www.saverilawfirm.com/our-cases/github-copilot-intellectual-property-litigation}} For example, in a landmark 2024 ruling, the Court of Appeal of Paris ordered Orange S.A., a French telecommunications company, to pay over €900{,}000 in damages for GPL violations.\footnote{\url{https://www.dlapiper.com/en/insights/publications/2024/03/wakeup-call-for-open-source-users-french-court-awards-damages-for-gpl-violations}} Additionally, an ongoing lawsuit against television manufacturer Vizio aims to compel the release of its proprietary source code to the public, which allegedly uses GPL code.\footnote{\href{https://www.dlapiper.com/en/insights/publications/2024/01/sfc-v-vizio-survives-motion-for-summary-judgment-on-third-party-beneficiary-issue\#2}{SFC v. Vizio case}}
 These examples demonstrate the severe financial and reputational consequences of open-source license non-compliance. Moreover, relying on the closed-source nature of code to ignore open-source licensing issues is a risky strategy, as non-compliant code can be revealed by internal or external audits (e.g., by vendors or contractors), regulatory reviews, whistleblowers, disgruntled employees, or accidental breaches. As traditional copyright laws in the U.S. and countries around the world are not fully equipped to address the uniquely probabilistic nature of ACA-generated content, we expect lawsuits to continue to emerge globally in the future.

Closed-source ACAs by large silicon-valley firms now provide some level of IP indemnity to paid customers under specific settings,\footnote{\url{https://resources.github.com/learn/pathways/copilot/essentials/establishing-trust-in-using-github-copilot/}} but there are a plethora of other ACAs that may not offer any protection.\footnote{\url{https://www.reuters.com/legal/legalindustry/legal-primer-open-genai-models-2024-08-15}} The need for these indemnity protections is a sign of the gravity of the issue and a tacit acceptance that ACAs can generate code governed by share-alike licenses such as GPL or AGPL exposing users to significant risk {[}2, 3, 4{]}. However, indemnity protections are often subject to caveats that limit the ACA provider's liability and can be invalidated by routine software development tasks. It is no wonder that there is significant uncertainty among lawyers regarding how broad and useful they are, how they would apply in practice, and whether they will fully protect the customers against financial loss.\footnote{\url{https://katedowninglaw.com/2023/11/02/yes-github-finally-offered-to-indemnify-for-copilot-suggestions-but/}} Finally, there could be significant variations in how well they work outside of U.S. in countries with different laws and regulations. Thus, for IT managers globally, the challenge isn't just about successfully adopting these tools, it's also about doing so without exposing their companies to legal and compliance nightmares {[}2{]}.

While large software firms have access to ACAs trained on proprietary code developed in highly regulated sandboxes, most small to medium-sized firms across the globe do not have the resources for such luxuries. Worse, they may be unaware of ACA-generated compliance risks. What proactive steps can IT managers in such firms take to prevent ACA-generated compliance challenges, and how should they respond if issues arise? We present a framework, \emph{DevLicOps}, that seeks to provide guidance by outlining (1) preventative governance measures, (2) incident response strategies (triage), and (3) tradeoff factors involved in making informed choices. DevLicOps integrates with the traditional software development life cycle (SDLC) model to prevent ACA-generated licensing conflicts from occurring in the first place. Triage consists of steps managers may consider when such conflicts are found in compiled code that has already been released. Finally, because not all organizations are alike, the framework also provides flexibility in how managers may adopt the framework depending on their risk tolerance and the code module being developed.

\section{Preventing ACA-generated license conflicts using DevLicOps}\label{preventing-aca-generated-license-conflicts-using-devlicops}

ACAs are poised to radically affect various stages of SDLC including requirements gathering and analysis, design, development, testing, deployment, and maintenance.\footnote{\url{https://kpmg.com/kpmg-us/content/dam/kpmg/pdf/2023/KPMG-GenAI-and-SDLC.pdf}} DevLicOps, which stands for development, licensing, and operations, integrates license compliance in all phases of SDLC where ACAs can introduce risks. DevLicOps can be seen as layer separate from DevSecOps, which focuses primarily on security vulnerabilities. DevLicOps promotes early planning and shared responsibility among all individuals involved in software production and deployment. It is designed to integrate with existing DevOps practices. It complements the DevOps workflow by incorporating license compliance checks and risk management steps for ACA-generated code without disrupting speed, collaboration, and efficiency in software development. DevLicOps is not a one-size-fits-all approach and managers have the flexibility to determine the degree of adherence to the framework depending on their context. We outline DevLicOps steps in the SDLC model below:

\subsection{Before coding begins - DevLicOps in planning and design:}\label{before-coding-begins---devlicops-in-planning-and-design}

As Aristotle famously declared, ``Well begun is half done.'' Setting the tone at the beginning of the project ensures that policies, roles, and compliance strategies are in place before coding begins to mitigate risks associated with ACA-use {[}1{]}.

\begin{enumerate}
\def\labelenumi{\arabic{enumi}.}
\item
  Establish clear ACA-use policies based on risk tolerance early
\end{enumerate}

An overarching ACA-use policy that aligns with the firm's strategic, business objectives and risk tolerance should be defined early in the project lifecycle. At the strategic level, risk appetite statements should include a consideration of how ACA-generated licensing compliance risks may jeopardize company goals.\footnote{\url{https://www.theirm.org/media/6878/0926-irm-risk-appetite-12-10-17-v2.pdf}} At the operational level, risk averseness to specific types of OSS licenses should be considered. This includes clearly identifying acceptable vs.\ non-acceptable code matching open-source licenses (e.g., AGPL in the case of Google)\footnote{\url{https://opensource.google/documentation/reference/using/agpl-policy}}, inter-license compatibility (irreconcilable requirements), as well as specific modules where ACA-use will be allowed based on their criticality. Core modules may need stricter policies than auxiliary modules, as we discuss below. The primary risk lies in ACA-generated code that matches, depends on, or is derived from copyleft licenses, which could mandate the entire source code to be released under the same license {[}2, 3{]}. To avoid this, clear policies should be set for tagging ACA-generated code for review and distinguish it from proprietary code. This may include policies for ACA-usage analytics to give a sense of how much code is ACA-generated vs.\ self-developed by monitoring adoption rates of code suggestions, total acceptances, active daily users etc. A thorough review of indemnity clauses and configuration settings is crucial for determining the level of protection. Managers should assess the need for additional indemnity protections for unexpected outcomes and verify whether clients permit ACA usage in case of contracted projects to ensure compliance with region-specific laws and industry regulations.

\begin{enumerate}
\def\labelenumi{\arabic{enumi}.}
\setcounter{enumi}{1}
\item
  Select the right ACA tool(s)
\end{enumerate}

Emerging research shows that regurgitation rates increase with ACA complexity and size (overparameterization), duplicates in training data (e.g., number of times a model is exposed to copyright code), recency of training data, number of tokens used in the prompt, and number of prompts by the user {[}4, 7{]}. The rates also differ across foundational model architectures (e.g., BERT, GPT, or Encoder-Decoder families). As ACAs become more complex in search of emergent properties, regurgitation risks will likely increase in the future {[}8{]}. Regurgitation by ACAs trained on copyright data can put users to significant risk. Thus, selecting and using ACAs wisely is critical to managing licensing concerns. Key factors to consider include the foundation model, size, training data quality and transparency, performance, cost, user-friendliness and compatibility (e.g., integration with IDEs), open vs.\ closed access, contractual requirements, and most critically, code filtering and indemnity protections.

\begin{enumerate}
\def\labelenumi{\arabic{enumi}.}
\setcounter{enumi}{2}
\item
  Understand the implications of ``output'' indemnity protections (or the absence thereof) on coding practices
\end{enumerate}

Understanding output indemnity protections, or a lack thereof, is critical from a compliance perspective because it determines how ACA-generated code may be used and significantly impact developers' workflows. Paid versions of closed-access ACAs by large providers provide content filters and indemnity protections for generated code output but these come with many exclusions. While these are steps taken to instill confidence in users, in our reading we found the exclusions to be so limiting that for most firms it may be best to assume that there are no protections at all. For instance, almost all indemnity clauses we read excluded coverage for scenarios where the output is modified, transformed, or used in combination with products or services from other providers. These coding practices are simply unavoidable during software production. Developers often iterate between coding and testing, relying on several tools to modify and update code. In fact, a recent survey found that 67\% of developers now report spending more time debugging, modifying, and refining buggy ACA-generated code than writing code themselves.\footnote{\url{https://www.harness.io/state-of-software-delivery}} Even worse, if the generated code based on the user’s input infringes on copyright, the customer may be required to provide indemnity protections to the provider.\footnote{\url{https://www.morganlewis.com/blogs/sourcingatmorganlewis/2024/06/contract-corner-ensuring-ip-provisions-are-fit-for-genai}} In other words, providing the ACA with a detailed contextual prompt to generate code that could be simply ``plugged in'' without modifications could also be risky.

Other exclusion criteria place responsibility on customers and their end-users to recognize or anticipate potential infringement in the generated output and are so broadly defined that they may be nearly impossible to refute in the event of claims. All large ACA providers have very similar exclusions and assign responsibility on users to conduct code and IP scanning on their end. Fine-tuning a model for contextual use could also invalidate indemnity protections.\footnote{\url{https://www.law.com/legaltechnews/2024/02/09/gen-ai-providers-offer-ip-indemnity-heres-why-its-not-fool-proof/}} Moreover, the reliability of content filters remains uncertain, as they have been shown to be easily circumvented {[}9{]}. Enabling the content filters could negatively impact the quality of code suggestions by forcing ACAs to be overly cautious and provide less creative and more generic outputs. This creates a catch-22, as such code cannot be modified to make it usable without breaking indemnity.

Open-source ACAs are free to use and offer greater flexibility for fine-tuning but are provided ``as is'' without any indemnity protections. While the current definition of open-source AI requires open code, open weights, and transparency about training data, it allows models to be trained on proprietary code and code governed by restrictive licenses.\footnote{\url{https://thenewstack.io/the-case-against-osis-open-source-ai-definition/}} Open-source ACAs trained on proprietary code may also regurgitate copyrighted material. To avoid copyright concerns associated with restrictive licenses and proprietary code, some ACAs, like StarCoder/StarCoderBase\footnote{\url{https://huggingface.co/bigcode/starcoderbase}} and InCoder,\footnote{\url{https://sites.google.com/view/incoder-code-models}} are trained exclusively on filtered GitHub code under permissive licenses. While StarCoder is a standard left-to-right code completion tool, InCoder is an infilling-capable code completion model that is designed to fill gaps by leveraging both preceding and following code and may be more useful for repeated editing tasks. However, because they are trained preferentially on code, such tools are less effective with text-based prompts (e.g., ``write a function that\ldots''). They may also lack integration with IDEs. Nonetheless, they can be selectively employed to replace flagged copyrighted code generated by proprietary ACAs. If feasible, customized ACAs fully trained on proprietary code or selected open-source projects with permissive licenses (e.g., BSD-3, Apache, or MIT) or open datasets such as BigCode Stack can be utilized. Alternatively, fine-tuning pre-trained ACAs on selected proprietary code or using Retrieval-Augmented Generation (RAG) may help minimize the risk of regurgitating copyrighted material. However, it is important to note that fine-tuning may invalidate indemnity protections.

\begin{enumerate}
\def\labelenumi{\arabic{enumi}.}
\setcounter{enumi}{3}
\item
  Implement ``indemnity preserving practices'' (IPPs) if indemnity protections are to be relied upon
\end{enumerate}

If vendor-provided indemnity protections are to be relied upon, IT managers must not only understand the exclusions but also strictly enforce what we term as ``indemnity preserving practices'' (IPPs). This is especially critical for firms with low risk tolerance to ensure their coding practices do not break indemnity protections. Because indemnity exclusions are broadly worded, vary among major vendors, and may evolve with time, we recommend IT managers to select a single ACA tool that best fits their needs and tailor their IPPs according to its specific terms. Legal advice is strongly recommended during this process. However, we caution that IPPs may impose significant adjustments and constraints on coding workflows, which many managers may find challenging to overcome.

In the strictest sense, IPPs should include at a minimum: (1) ensuring no modifications or transformation to the ACA-generated code are made by developers, (2) avoiding the use of ACA-generated code in combination with products or services from other providers, (3) curbing shadow AI practices, where developers use non-approved ACAs, (4) maintaining accurate and complete code license citations, (5) ensuring that content filters and other safety features are not bypassed, (6) avoiding the input of copyright infringing code into the ACA to generate code, (7) preventing the use of ACA-generated code to conduct business, either by the firm or its downstream end-users, in ways that could violate trademarks or copyrights, and most critically, (8) assuming full responsibility for verifying whether ACA-generated code violates any copyright, as this burden is typically placed on the users by ACA-providers. Unless IPPs are strictly followed, managers should not rely on the security provided by indemnity protections. Furthermore, given that the fine prints of indemnity agreements may change with time, managers must remain vigilant and update IPPs as needed. Given that developers increasingly spend time debugging, refining, and modifying buggy ACA-generated code, IPPs may be implementable only in very restricted settings.

\begin{enumerate}
\def\labelenumi{\arabic{enumi}.}
\setcounter{enumi}{4}
\item
  Create a risk-aware culture through training
\end{enumerate}

A recent survey study of 574 developers found that only 68 (11.9\%) developers indicated that they had any formal training in copyright law and/or the legal implications of using code generation models. Of these, only 16 noted they received training from their employer {[}10{]}. This lack of awareness creates a serious risk for companies. Companies should provide developer training on license compliance and risks with ACA-generated code through training programs and videos. This includes covering key topics such as ACA-use policies, open-source license compliance, attribution requirements, acceptable modification and distribution, copyright laws (such as fair use in the U.S.), contractual requirements, and use of tools for license compliance. Furthermore, periodic training sessions should be conducted to keep developers updated about evolving compliance risks and industry best practices. This is especially relevant for less-experienced developers, who are less likely to have prior exposure to such issues but may be most reliant on ACAs. Such training should discuss IPPs and specific coding activities (e.g., modifying or editing code) that could invalidate any indemnity protections, if they are being relied upon.

\begin{enumerate}
\def\labelenumi{\arabic{enumi}.}
\setcounter{enumi}{5}
\item
  Define compliance roles for license audit teams
\end{enumerate}

Assign clear roles and responsibilities for license compliance by specifying who will monitor ACA-generated code, how often checks will occur, and which tools will be used. Establish a well-trained compliance audit team to oversee adherence to ACA-use policies, resolve conflicts, and ensure smooth communication between developers, managers, and legal experts. Define escalation procedures to address violations effectively.

\begin{enumerate}
\def\labelenumi{\arabic{enumi}.}
\setcounter{enumi}{6}
\item
  Create a roadmap for rapid adaptation
\end{enumerate}

The AI landscape is very dynamic with constant changes in technological advancements and industry trends. This necessitates creating a roadmap for rapidly adapting to changes. At a minimum, this roadmap should include (1) continuously monitoring technological and regulatory trends to anticipate and mitigate compliance risks, (2) updating organizational compliance policies based on lessons learned from practical experience, audits, and external information, (3) periodically checking for open-source licensing updates, as previously compliant code may become subject to new licensing terms, and new licenses may emerge, and (4) monitoring and adapting to changes in ACA-providers' indemnity and use policies.

\subsection{Once coding has begun - DevLicOps in production and development:}\label{once-coding-has-begun---devlicops-in-production-and-development}

This phase focuses on preventing ACA-generated licensing conflicts during the coding stages, the most vulnerable phase in SDLC. Catching the violations here can prevent downstream effects. Managers can implement the suggested steps to varying degrees while balancing compliance with development efficiency. Preventing infringing code from entering production will require creating lines of defense with automated tests, manual audits, and isolated third-party repositories to quarantine code.

\begin{enumerate}
\def\labelenumi{\arabic{enumi}.}
\item
  Configure content filters for indemnity compliance
\end{enumerate}

If using ACAs embedded in the IDEs, the first step is to enable copyright content filter settings according to the ACA provider's guidelines and document these configurations in IDE readme files and repository guidelines to ensure consistent compliance practices across teams. This is especially important if indemnity protections are to be relied upon. However, as discussed above, the effectiveness of content filters in preventing copyright violations is yet to be known. Further, content filters may degrade the output code quality.

\begin{enumerate}
\def\labelenumi{\arabic{enumi}.}
\setcounter{enumi}{1}
\item
  Implement automated compliance monitoring tools
\end{enumerate}

Perhaps the most critical step is to embed automated compliance monitoring through Software Composition Analysis (SCA) tools in the Continuous Integration/Continuous Deployment (CI/CD) pipelines. Automated scans can be triggered during the ``code push'' stage, where developers transfer code from their local branch to the remote development branch (e.g., \texttt{/dev}). Developers often interact with multiple ACAs and import code through various means, such as ACAs embedded in IDEs, plugins, or by copying and pasting code from external ACAs. These workflows involve substantial back-and-forth adjustments to ensure functionality. Developers could also rely on shadow AI practices. Thus, the push stage is the first reliable point to comprehensively check for license violations.

SCA tools specializing in OSS license detection, such as SCANOSS, FOSSA, and Black Duck, can seamlessly integrate into CI/CD pipelines. These tools can automatically scan ACA-generated code against predefined lists of acceptable and unacceptable licenses, and generate detailed license compliance reports that highlight violations, affected code lines, copy percentages, and associated risk levels based on company policy. At this stage, automated scripts can flag and block code pushes in the event of critical violations, requiring developers to either fix the issues or document exceptions. These exceptions become part of a virtual paper trail for ensuring accountability and traceability. Once the code passes automated tests, it can be merged into the main branch and made available for manual audits. The procedure for addressing flagged code may vary based on the following considerations:

\begin{itemize}
\item
  \textbf{Type of code used to train the ACA:} If the organization relies on ``safe'' ACAs, such as StarCoder or InCoder, which are trained exclusively on permissively licensed code, the flagged code may only require proper license attribution to remain compliant. However, in case of ACAs trained on unfiltered datasets, including code governed by restrictive licenses, the flagged code may need to be blocked, depending on the specific license detected. For example, if the flagged code matches GPL or AGPL, the developer may need to manually rewrite the code from scratch or use output from a safe ACA. Copyright laws typically focus on the ``heart'' or essence of the code, which means that merely altering a few lines of code to pass the automated tests is unlikely to be sufficient. The extent of transformation required depends heavily on the context, as there is often a ``blurry line between ideas and expression in code'' {[}3{]}. Copyright protections extend beyond superficial changes, and the fair use doctrine generally requires both low-level transformations (e.g., reducing n-gram overlap) and higher-level modifications (e.g., altering non-generic code structures) {[}3{]}. Depending on the amount of overlap found, substantial modifications at every scale of code may be required. Since developers are unlikely to be copyright subject experts, these modifications must be documented in detail for later manual audits by trained teams.
\item
  \textbf{Organizational risk tolerance and implementation of IPPs:} In case of low risk tolerance organizations or ones implementing IPPs strictly, flagged code should not be modified at all. Instead, the flagged code must be discarded entirely and replaced with code developed in full compliance with company policies and ACA provider's indemnity clauses.
\item
  \textbf{Code module under development:} As discussed below, copyrighted code poses the greatest threat to core business logic modules and a comparatively lower risk to auxiliary and support modules. Flagged ACA-generated code within core module development represents the highest level of organizational risk and should be blocked entirely until a thorough review can be conducted. In contrast, flagged violations in experimental or auxiliary module development may be assigned a lower priority initially. However, if left unresolved, these violations could still pose significant issues if they are later incorporated into production code.
\end{itemize}

\begin{enumerate}
\def\labelenumi{\arabic{enumi}.}
\item
  Periodic manual audits by trained experts
\end{enumerate}

Manual audits by teams trained in copyright and compliance regulations can ensure that guardrails of human judgement and nuanced copyright interpretation are preserved. Unlike automated tests, manual audits can account for more nuanced copyright and compliance policies. These audits can verify that developers’ modifications to ACA-generated code are not just superficial attempts to bypass automated checks but are instead aligned with compliance and organizational standards. Audit teams should have the authority to overrule or block any commits that fail compliance requirements from entering production code. The frequency of these audits should be managed carefully so as not to overwhelm the team by large portions of the codebase. Audit teams can prioritize reviewing flagged code identified by considering the type of module under development and the extent to which flagged code may pose a copyright risk.

All flagged code commits should be manually audited, especially those flagged multiple times or deemed high-risk by SCA tools. Audit teams can query repositories to audit blocked commits, access the associated SCA reports, and review changes made by developers. Audit reports may include a list of ``items,'' each with a flagged commit number, its corresponding fixed commit number, SCA license compliance report or a link, record of corrective actions, and other relevant information such as timestamps and developer details. Audit teams can use this information to determine whether the changes comply with licensing requirements. Although manual audits are resource-intensive, they can be invaluable when performed by well-trained experts.

\begin{enumerate}
\def\labelenumi{\arabic{enumi}.}
\setcounter{enumi}{2}
\item
  Create a secure, isolated directory for third-party ACA code
\end{enumerate}

Code generated by third-party or non-approved ACAs should be stored in a separate directory until it is reviewed and cleared for use. For example, Google uses a ``//third\_party'' directory for all non-Google-owned code, including open-source code.\footnote{\url{https://opensource.google/documentation/reference/thirdparty}} This centralized location simplifies tracking all external code, licensing restrictions, third-party dependencies, and ensuring compliance before use.

\begin{enumerate}
\def\labelenumi{\arabic{enumi}.}
\setcounter{enumi}{3}
\item
  DevLicOps steps in testing and pre-deployment
\end{enumerate}

Incorporating compliance checks can be the last line of defense before its deployment. Without proper checks, non-compliant code may slip through the testing stages leaving a vulnerable product that puts the company at risk. Compliance checks should continue throughout the product lifecycle, particularly when newer versions are introduced. If new ACA-generated code is added during testing, additional automated SCA checks and manual audits should be conducted, as outlined above. Additionally, all older code should be checked for continued license compliance, as licenses may have changed since the last audit. The software bill of materials (SBOM) should also be updated to account for all licenses across the entire codebase. Finally, we recommend consulting with legal teams for a review in case of any concerns raised by audit teams before deployment.

\section{Responding to ACA-generated license conflicts post-deployment (Triage)}\label{responding-to-aca-generated-license-conflicts-post-deployment-triage}

Responding efficiently to licensing conflicts post-deployment requires categorizing the issues as high, medium, or low severity. High severity issues, such as copyleft violations affecting the core business logic, site reliability, or data pipelines, could imperil the entire business and necessitate immediate remedial actions. Once flagged, the offending code must be isolated, removed, and replaced with compliant code, either through custom-built solutions or by using safe ACA outputs. While these steps are ongoing, it is crucial to have a robust rollback plan to revert problematic changes and notify stakeholders to effectively manage downstream impacts. If the issue has occurred despite following ACA-provider indemnity requirements and IPPs, consult with the provider to understand how modifications can be made without invalidating indemnity and obtain written confirmation for any changes. If a resolution is not achievable, then consider filing an indemnity protection claim. Throughout this process legal consultation is necessary, but ethical considerations such as transparency with users and acknowledging accountability should not be overlooked.

Moderate severity issues might include missing permissive license attributions, which can often be resolved by simply referencing the original work to ensure compliance. In other cases, copyleft license conflicts in auxiliary or non-critical helper modules can be addressed by making the code copyleft-compliant and publicly available without jeopardizing the organization’s intellectual property. Low severity issues could include minor permissive license discrepancies (e.g., updates from MIT to Apache licenses) or other inaccuracies in attribution and can typically be addressed in the next development cycle.

\begin{figure}[H]
  \centering
  \includegraphics[width=\linewidth]{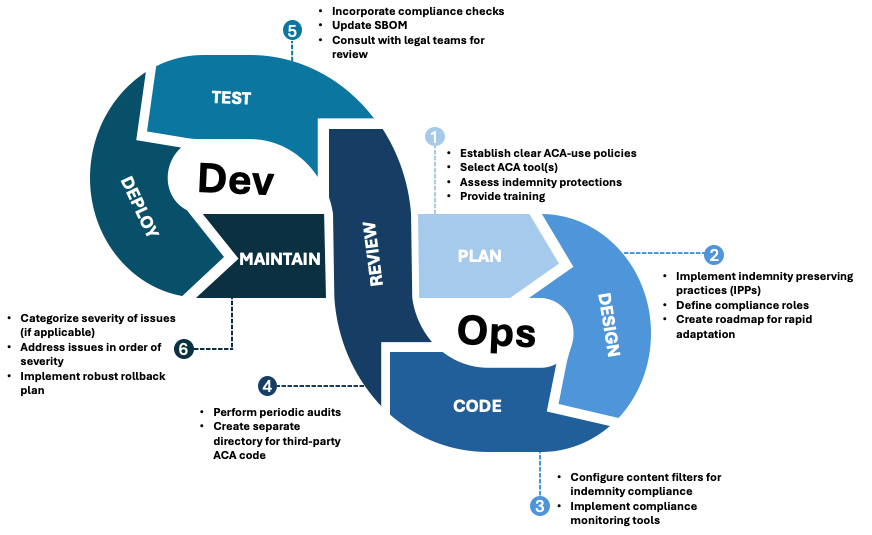}
  \caption{The DevLicOps Framework}
  \label{fig:devlicops}
\end{figure}

\section{Understanding the tradeoffs in applying DevLicOps}\label{understanding-the-tradeoffs-in-applying-devlicops}

ACAs will reshape the role of IT managers to that of curators who must balance the performance gains associated with ACA use with a thorough evaluation of associated risks. They will need to navigate the complex interplay of legal, reputational, and financial risks their organizations may face and make risk-informed decisions. This requires identifying, assessing, monitoring, and mitigating license compliance risks introduced by ACA-generated code.

The specific DevLicOps steps adopted will depend on the organization's risk tolerance. In risk-averse sectors such as financial services, defense, and healthcare, with ample resources, managers may prioritize stringent safeguards such as comprehensive manual audits. Conversely, in resource-constrained startups with a higher risk appetite they may adopt a more agile strategy and focus on selective oversight and reactive measures. Table 1 discusses some of the tradeoffs involved.

\noindent\textbf{Table 1. Tradeoffs}

\begin{center}
\begingroup\small
\begin{longtable}[]{@{}
  >{\raggedright\arraybackslash}p{(\linewidth - 6\tabcolsep) * \real{0.2312}}
  >{\raggedright\arraybackslash}p{(\linewidth - 6\tabcolsep) * \real{0.1951}}
  >{\raggedright\arraybackslash}p{(\linewidth - 6\tabcolsep) * \real{0.2098}}
  >{\raggedright\arraybackslash}p{(\linewidth - 6\tabcolsep) * \real{0.3639}}@{}}
\toprule
Step & Upfront investment & Continuous implementation effort & Factors impacting effectiveness \\
\midrule
\endhead
Establishing ACA policy on license compliance & Low & Low & Effectiveness depends on the comprehensiveness of the policy and how consistently it is implemented. \\
Educating developers (risk-aware culture) & Medium & Medium & Effectiveness depends on the quality of training and the degree to which developers apply the knowledge. \\
Creating secure software environment (Sandbox, third-party directories) & Medium & Medium & A secure environment minimizes contamination risks by allowing audits before code moves to production. \\
Using indemnified ACAs vs.\ non-indemnified (``as is'') ACAs & Medium & Low & The effectiveness of content filters in identifying copyright violations is uncertain and can be easily circumvented {[}9{]}. Additionally, filter accuracy varies across providers. \\
Deploying customized ACAs fully trained on self-owned proprietary code & Very High & Medium & In theory, ACAs trained exclusively on proprietary code should not generate outputs with licensing conflicts. \\
Deploying customized ACAs partly trained on self-owned proprietary code (e.g., RAGs or selected projects) & Medium & Medium & Partially trained ACAs reduce the likelihood of licensing violations but do not eliminate them entirely. \\
Implementing automated compliance in CI/CD pipelines (e.g., using FOSSA) & Medium & Medium & Effectiveness depends on the thoroughness of controls and developer adherence. It also relies on the efficiency of SCA tools like FOSSA. For enhanced protection, multiple license detection tools should be used. \\
Manual audits & High & High & Effectiveness depends on the diligence and expertise of the license auditing team. \\
Maintaining a legal team & High & High & Effectiveness depends on the team’s level of diligence, involvement, and expertise in open-source licensing. \\
\bottomrule
\end{longtable}
\endgroup
\end{center}

Additionally, risk tolerance will vary by the type of module under development (Table 2 below). Core or proprietary modules, which are integral to a company's competitive advantage, represent higher stakes, as any licensing violations could compromise entire business by exposing proprietary code to restrictive licenses. In contrast, auxiliary or non-critical modules carry lower risks and allow firms to adopt less stringent measures while maintaining efficiency. Understanding these nuances can help IT managers tailor their approach according to their context.

\noindent\textbf{Table 2. Module-Specific Risk Severity and Mitigation for ACA-Generated Code}

\begin{center}
\begingroup\small
\begin{longtable}[]{@{}
  >{\raggedright\arraybackslash}p{(\linewidth - 6\tabcolsep) * \real{0.1534}}
  >{\raggedright\arraybackslash}p{(\linewidth - 6\tabcolsep) * \real{0.1146}}
  >{\raggedright\arraybackslash}p{(\linewidth - 6\tabcolsep) * \real{0.3242}}
  >{\raggedright\arraybackslash}p{(\linewidth - 6\tabcolsep) * \real{0.4079}}@{}}
\toprule
Module type & Severity & Impact of licensing violation & Suggested DevLicOps Steps \\
\midrule
\endhead
Core, Proprietary Business Logic & Very High & Could expose proprietary core code to forced distribution under restrictive licenses putting entire business to risk. & Proactively engage with legal experts to create an ACA-specific policy which considers licensing risks. Consider conducting all DevLicOps steps including code quarantines for ACA-generated code, use of custom-trained ACAs (if possible), and frequent automated and manual audits. Be vigilant for open-source dependencies and changes in licensing over time. \\
Site Reliability, Data Pipelines & High & Critical system functionality and data processing ability may be impacted. & Consider conducting most DevLicOps steps. Depending on the amount of coding needed, periodic manual audits may suffice. \\
User interface & Medium & Front-end elements may be impacted requiring design changes that could impact customer experience. & Use pre-approved, indemnified ACA tools. Monitor ACA use and track ACA-generated code. Conduct automated license compliance for frontend libraries and frameworks. \\
APIs & Medium & Interfaces and integration points may be affected disrupting interconnectivity. & Use pre-approved, indemnified ACA tools. Monitor ACA use and track ACA-generated code. Conduct automated license compliance. Review API contracts \& implement ACA-policies for compliance. \\
Prototype & Low & Lower impact of violation in the short-term as code is experimental. However, violations may have significant impact later if code goes to production. & Monitor and tag ACA-generated code. Flag ACA-generated code for review if reused in production. Educate teams on prototype-to-production transitions. Focus on efficiency in code development over exhaustive audits. \\
Auxiliary (support) & Low & Violation in supportive modules could cause short-term, limited impact on some functionality. & Monitor and tag ACA-generated code. Use automated scans. Define clear guidelines for ACA use in auxiliary modules. Focus on efficiency in code development over exhaustive audits. \\
\bottomrule
\end{longtable}
\endgroup
\end{center}

\section{Conclusion}\label{conclusion}

The purpose of our article is to: (1) raise awareness among IT managers, particularly in small to medium-sized firms, regarding license compliance risks with ACA-generated code, (2) provide guiding steps for implementing best practices that protect their organizations, and (3) suggest possible course of action when problems have occurred. The steps we propose are informed by best practices at large firms, insights from discussions with IT professionals, and our own academic and professional experiences. While not exhaustive, these steps offer a starting point and will require refinement as ACAs continue to evolve.

In informal conversations, we found significant uncertainty among upper management regarding license compliance issues with ACA-generated code. As a precaution, some are considering limiting or fully avoiding ACA-usage. However, this approach may not be sustainable from a long-term competitive standpoint. In the near term, indemnity protections offered by major ACA providers are a welcome step toward building customer confidence. However, we caution that such protections are not comprehensive and should not be depended upon. Much like life insurance, which covers many scenarios but excludes high-risk activities, indemnity protections include significant limitations and exclusions. Even if indemnity protections are ultimately activated, their actual value may be limited in many cases, and recovering losses can take an extended period leaving the firm vulnerable in the interim.

Given that global copyright laws are playing catch up with rapid technological advancements and that indemnity protections vary across ACA providers, firms must remain vigilant about compliance risks when using ACAs. As a 2023 \emph{Wired} article aptly observes, ``{[}the{]} Generative AI copyright fight is just getting started.''\footnote{\url{https://www.wired.com/story/livewired-generative-ai-copyright/}} While the initial wave of lawsuits has targeted ACA providers, the next wave could very well focus on downstream ACA users. In the U.S., fair use doctrine has protected ACA providers building models using copyright code. However, fair use may not protect the users of the output of those models if it matches copyright content {[}3, 4{]}. Such lawsuits may originate from competitors, enforcement organizations, and copyleft attribution trolls and not necessarily from open-source code creators themselves.\textsuperscript{15,} \footnote{\url{https://www.techdirt.com/2021/12/20/beware-copyleft-trolls/}} Regardless of the threat of lawsuits, proactively addressing ACA-generated licensing issues and providing due credit are integral to responsible AI and ethical software development practices. Thus, the safest and most responsible guiding principle for IT managers may well be: \emph{``If we deployed it, we own up to it.''}

\section*{REFERENCES}
\begin{enumerate}
\item I. G. Cohen, T. Evgeniou, and M. Husovec. 2023. Navigating the new risks and regulatory challenges of GenAI. Harvard Business Review. Retrieved from \url{https://hbr.org/2023/11/navigating-the-new-risks-and-regulatory-challenges-of-genai}
\item A. Buttell. 2025. Protect your code against licensing risks. Communications of the ACM (Jul.\ 21, 2025). Retrieved from \url{https://github.blog/news-insights/research/survey-ai-wave-grows/}
\item P. Henderson, X. Li, D. Jurafsky, T. Hashimoto, M. A. Lemley, and P. Liang. 2023. Foundation models and fair use. \emph{Journal of Machine Learning Research} 24, 400 (2023), 1--79.
\item A. Al-Kaswan, M. Izadi, and A. Van Deursen. 2024. Traces of memorisation in large language models for code. In \emph{Proceedings of the IEEE/ACM 46th International Conference on Software Engineering (ICSE 2024)}. IEEE, 1--12.
\item N. Carlini, D. Ippolito, M. Jagielski, K. Lee, F. Tramèr, and C. Zhang. 2022. Quantifying memorization across neural language models. \emph{arXiv preprint} arXiv:2202.07646.
\item T. Chu, Z. Song, and C. Yang. 2024. How to protect copyright data in optimization of large language models? In \emph{Proceedings of the AAAI Conference on Artificial Intelligence} 38, 16, 17871--17879.
\item V. Smith, A. S. Shamsabadi, C. Ashurst, and A. Weller. 2023. Identifying and mitigating privacy risks stemming from language models: A survey. \emph{arXiv preprint} arXiv:2310.01424.
\item J. Katzy, R. Popescu, A. Van Deursen, and M. Izadi. 2024. An exploratory investigation into code license infringements in large language model training datasets. In \emph{Proceedings of the 2024 IEEE/ACM First International Conference on AI Foundation Models and Software Engineering (FMS'24)}. IEEE, 74--85.
\item D. Ippolito, F. Tramèr, M. Nasr, C. Zhang, M. Jagielski, K. Lee, and N. Carlini. 2022. Preventing verbatim memorization in language models gives a false sense of privacy. \emph{arXiv preprint} arXiv:2210.17546.
\item T. Stalnaker, N. Wintersgill, O. Chaparro, L. A. Heymann, M. Di Penta, D. M. German, and D. Poshyvanyk. 2024. Developer perspectives on licensing and copyright issues arising from generative AI for coding. \emph{arXiv preprint} arXiv:2411.10877.
\end{enumerate}

\end{document}